\documentstyle[seceq,wrapft,epsf]{ptptex}

\title{
Relaxational Modes and Aging in the Glauber Dynamics of\\
the Sherrington-Kirkpatrick Model}

\author{
Hajime {\sc Yoshino}\footnote{E-mail address  yhajime@ginnan.issp.u-tokyo.ac.jp},
Koji {\sc Hukushima},
and Hajime {\sc Takayama}
}

\inst{
Institute for Solid State Physics, Univ. of Toyko,\\
7-22-1 Roppongi, Minato-ku, Tokyo 106
}

\recdate{
\today
}

\newcommand{\disp}{\displaystyle}
\newcommand{\be}{\begin{equation}}
\newcommand{\ee}{\end{equation}}
\newcommand{\ba}{\begin{array}}
\newcommand{\ea}{\end{array}}
\newcommand{\ben}{\begin{eqnarray}}
\newcommand{\een}{\end{eqnarray}}
\newcommand{\bc}{\begin{center}}
\newcommand{\ec}{\end{center}}
\newcommand{\eq}[1]{(\ref{#1})}
\newcommand{\tw}{t_{\rm w}}
\newcommand{\eql}[2]{\begin{equation}\label{#1} #2 \end{equation} }

\abst{
The relaxational modes and aging of the Glauber dynamics of
the mean-field spin-glass model, SK model,
are studied by a numerical diagonalization technique and Monte Carlo
simulations.
We found that the aging process of the model is understood as 
hierarchical growth of quasi-equilibrium domain in the phase-space.
}

\begin{document}

\maketitle

\section{Introduction}

The fascinating features of aging effects of spin-glasses 
revealed by experiments \cite{experiment} have arose much theoretical interest
in recent years, 
which includes various phenomenologies \cite{B92}\cite{h-diff},
analytic predictions \cite{CK}  and numerical 
simulations.\cite{CK-sim}\cite{Bal}
In order to seek for concrete information to provide
theoretical base,
we study the relaxational modes of the 
Glauber dynamics of the mean-field spin-glass model, 
namely Sherrington-Kirkpatrick (SK)
model \cite{SK} of finite sizes at temperatures below $T_{\rm c}$
(spin-glass transition temperature).
By numerically diagonalizing the transition matrix
of small system sizes, we obtain a spectrum of relaxational modes
and analyze their properties to get insight into the mechanism
of the aging process. We also discuss the data of Monte Carlo 
simulations we have performed on larger system sizes.


\section{Model}

We study the SK model, whose Hamiltonian is given as,
\eql{sk}{{\cal H}=\sum_{ij} J_{ij}\sigma_{i}\sigma_{j}
-h \sum_{i}\sigma_{i}.}
where $\sigma_{i}$'s are Ising spins,
$\{J_{ij}\}$ are independent random Gaussian variables with zero 
mean and the variance $(N-1)^{-1}$ and $h$ is the external magnetic field.

We denote the probability that a spin configuration 
$\{\sigma^{\alpha}_{i}\}$ is realized at time $t$ as $p_{\alpha}(t)$. 
The time evolution of the probability
${\bf p}(t)=(p_{1}(t),p_{2}(t),\ldots,p_{2^N}(t))$, 
is detemined by the master equation,
\eql{Meq}{\disp\frac{d}{dt}{\bf p}(t) = -{\bf \Gamma} {\bf p}(t)
\mbox{\hspace*{1cm} with $-{\bf \Gamma}_{\alpha\beta}={\bf W}_{\alpha\beta}
-\delta_{\alpha\beta}
\sum_{\gamma}{\bf W}_{\gamma\beta}$}.}
The transition probability ${\bf W}_{\alpha\beta}$ is chosen as
the heat-bath type,
${\bf W}_{\alpha\beta}=\frac{1}{2}(1-\sigma_{i}^{\beta}
\tanh(T^{-1} (\sum_{j} J_{ij}\sigma_{j}^{\beta}+h)))$.
The unit of time corresponds with 1 Monte Carlo step/spin (MCS)
used in the standard Monte Carlo simulations. 

As usually done, it is convenient to introduce 
a matrix $\tilde\Gamma_{\alpha\beta}$
defined through the similarity transformation
${\bf \Gamma}_{\alpha\beta}
={\bf P}_{\rm eq}^{-1/2}{\bf \tilde\Gamma_{\alpha\beta}}
{\bf P}_{\rm eq}^{1/2}$.
The matrix $\tilde\Gamma_{\alpha\beta}$ is a real-symmetric matrix
and can be diagonalized as
${\bf \tilde\Gamma_{\alpha\beta}}=
\sum_{l=1}^{2^N}u_{\alpha}(l)z(l)u_{\beta}(l)$
where $u_{\alpha}(l)$ is the $\alpha$-th element of
the ortho-normal eigen-vector corresponding to the
eigen-valuse $z(l)$. We label the eigen-values in the 
order $0=z(1) < z(2) < \ldots z(2^N)$. The first mode 
corrsponds with the equilibrium distribution
and the eigen-vector is known exactly as 
$u_{\alpha}(1)=\sqrt{ p^{\rm eq}_{\alpha}}$.
In our analysis, we perform numerical diagnalization 
(Householder method) to obtain the
whole eigen-modes of system sizes $N=8,\ldots,12$. 

Once the eigen-modes are obtained, ${\bf G}_{\alpha\beta}(t)$,
the transition probability
to go from $\beta$ to $\alpha$ in time $t$, can be wirtten as
\eql{propagator}{
{\bf G}_{\alpha\beta}(t)=p^{\rm eq}_{\alpha}\sum_{l=1}^{2^N}
r_{\alpha}(l)\exp(-z(l)t)r_{\beta}(l)
\mbox{\hspace*{1cm} with 
$r_{\alpha}(l)\equiv u_{\alpha}(l)/\sqrt{ p^{\rm eq}_{\alpha}}$}.}

\section{Analysis}
\subsection{Aging after Temperature Quench}

Now we consider aging process after rapid temeprature quench
from the infinitely high temperature down to a temperature below $T_{\rm c}$.
For this purpose, we choose the initial condition for the master
equation as $p_{\alpha}(0)=1/2^{N}$, which means we choose the initial
spin configuration at random. Then the system ages as
\eql{aging}{
p_{\alpha}(\tw)=\sum_{\beta}{\bf G}_{\alpha \beta}(\tw)p_{\beta}(0).
}
by the propagator ${\bf G}_{\alpha \beta}(\tw)$ of a certain 
temperature $ T < T_{\rm c}$.

In order to measure the extent to which the system approachs the equilibrium
with a given $\tw$,
it is convenient to introduce the following {\it indicator of aging},
\eql{indicator}{
R_{\alpha}(\tw)\equiv  p_{\alpha}(\tw)/p^{\rm eq}_{\alpha}
=\sum_{l}r_{\alpha}(l)\exp(-z(l)\tw)\sum_{\beta}r_{\beta}(l)p_{\beta}(0)
.}

In Fig.~1 we plot an example of the evolution of the indicators
at spin configurations of different energy minima. 
Suppose that the indicators of
a pair of spin-configurations, say  $\alpha$ and $\beta$, satisfy the
equality with certain characteristic time $t^{*}_{\alpha \beta}$;
\eql{equality1}{
R_{\alpha}(\tw)=R_{\beta}(\tw) 
\mbox{\hspace*{2cm}for $\tw \gg t^{*}_{\alpha \beta}$}.
}
Then we say that the two configurations 
 are in {\it quasi-equilibrium},
because the relative probability that they are realized at 
$\tw \gg t^{*}_{\alpha \beta}$ is the same as in the true equilibrium.
In terms of eigen-modes, the above equality is equlivalent to the
following one: with $l^{*}_{\alpha\beta}$ defined such that
with $z(l^{*}_{\alpha\beta})\simeq (t^{*}_{\alpha\beta})^{-1}$,
\eql{equality2}{
r_{\alpha}(l)=r_{\beta}(l)
\mbox{\hspace*{3cm}for $1\leq l \leq l^{*}_{\alpha\beta}$}.}

\begin{figure}[htb]
  \parbox{\halftext}{
    \epsfxsize=6cm \epsfysize=3.9cm
    \centerline{\epsfbox{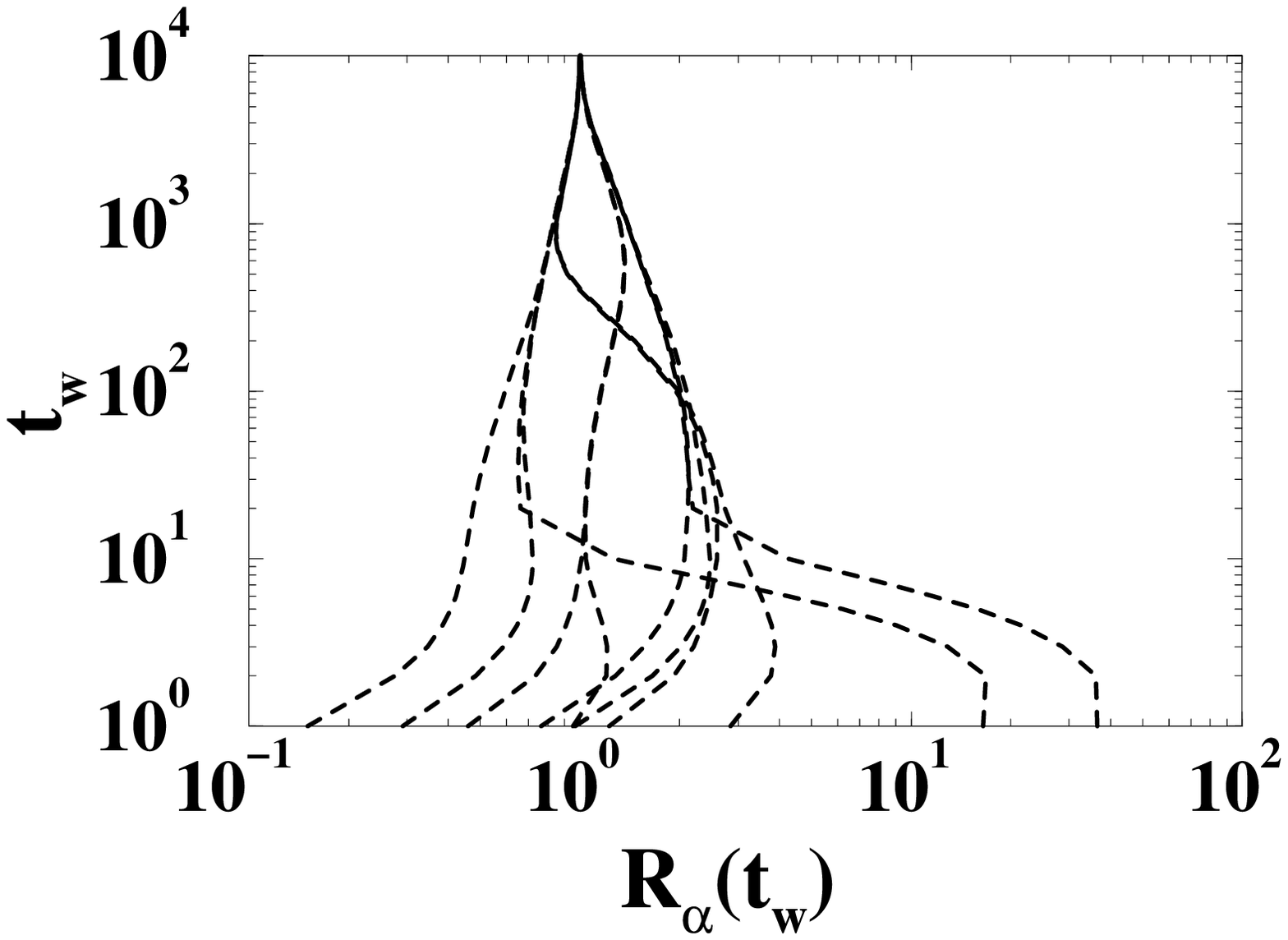}}
    \caption{The evolution of the indicator $R_{\alpha}(\tw)$
     for different energy minima $\alpha$. This is an example of
     $N=8$, $T=0.2$ and $h=0.05$.}}
    \hspace{8mm}
    \parbox{\halftext}{
      \epsfxsize=6cm \epsfysize=3.5cm
      \centerline{\epsfbox{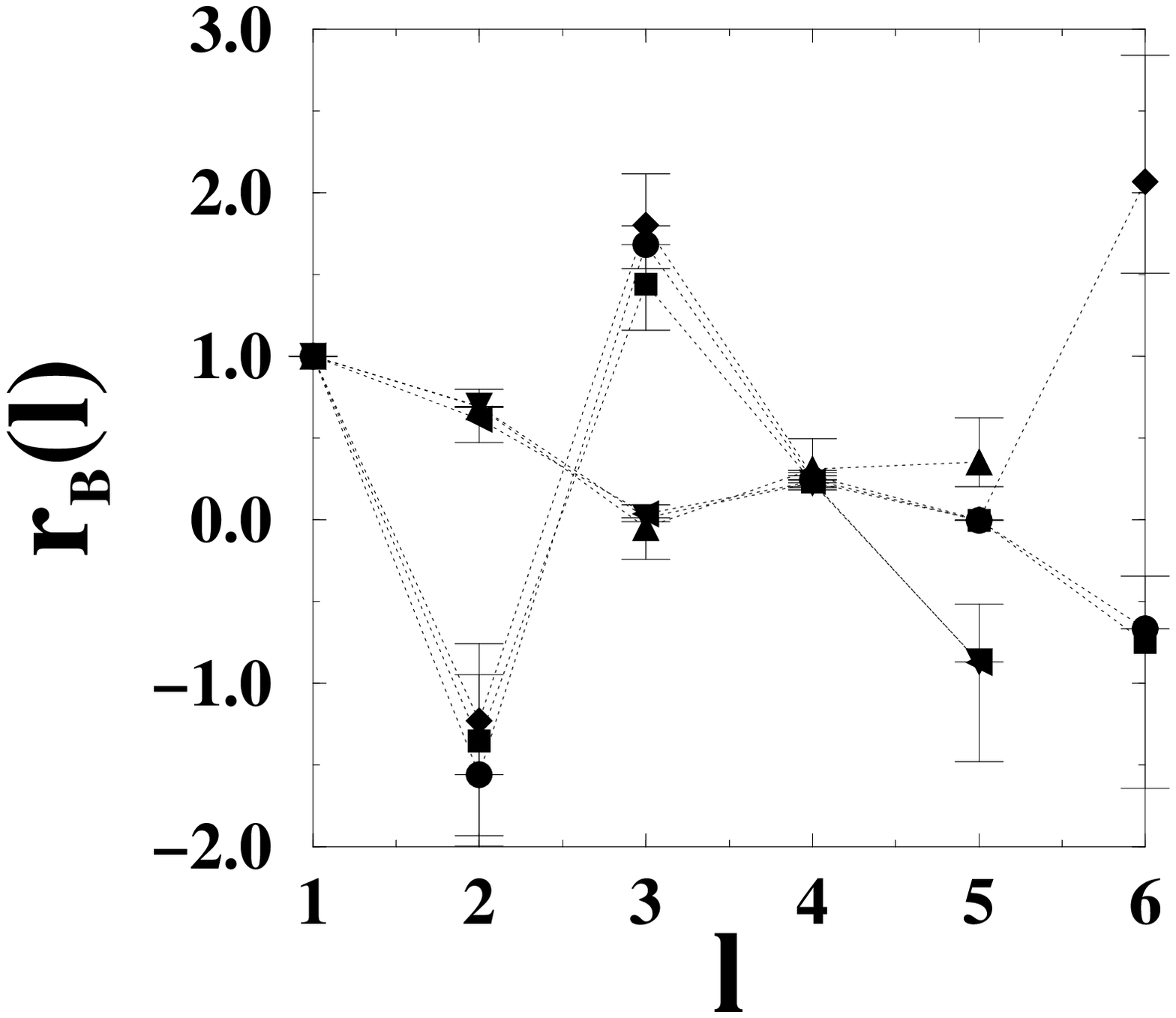}}
      \caption{The factor $\tilde{r}_{\cal B}(l)$ for different
        basins (different symbols) around the 6 energy minima of the
        system shown in Fig.~1.  
        The error bars represent the scatterings.}}
\end{figure}

\subsection{Slow Modes and 
Hierarchical Organization of Quasi-Equlibrium Domains}

Let us now introduce what we call {\it slow modes}.
There are a group of the eigen-modes
belonging to eigen-values $0 = z(1) < z(2) < \ldots z(l_{\rm th})$
with a threshould value $l_{\rm th}$ such that the maxima of their 
eigen-vectors locate at the elements corresponding 
to the energy minima.
We call this set of eigen-modes as {\it slow modes}.

For the slow modes, we have found that 
equality \eq{equality2} is satisfied 
within each {\it basin} ${\cal B}$, the neighbourhood
of an energy minimum in the phase-space, as 
\eql{tilde_r}{
r_{\alpha}(l) 
\simeq \tilde{r}_{\cal B}(l)e^{\delta_{\alpha}(l)}
\mbox{\hspace*{0.5cm}\small for $1 < l < l_{\rm th}$ and $\alpha\in  \cal{B}$} 
}
up to some small scattering factors
$e^{\delta_{\alpha}(l)}$.
In practice, we defined the basin
as a set of spin-configurations such that $T=0$ dynamics
starting from any one of it converge to the energy minimum 
with probability one. Then we obtained $\tilde{r}_{\cal B}(l)$ and
the scattering $e^{\delta_{\alpha}}$ by a $\chi^{2}$-fitting method.

The above result means that quasi-equilibrium within each basin is 
established at time scales specified by the slow modes.
For instance, we found that the relaxation time 
of the $l_{\rm th}$-th mode, which is $z(l_{\rm th})^{-1}$,
is about 6 MCS at $T=0.2$ independent of the system
sizes we studied ($N=8,\ldots,12$). 
On the other hand, the relaxation time of the second mode, which
is responsible for the time-reversal symmetry-breaking,
grows exponentially fast with $N$, as found previously \cite{YK}.

The apparent tree structure recognized in Fig.~1, is due to the
follwoing. For the factor $\tilde{r}_{{\cal B}}$ of a pair of basins,
say\/ ${\cal B}_{1}$ and ${\cal B}_{2}$, there is a characteristic
mode $l^{*}_{{\cal B}_{1}{\cal B}_{2}}$ such that the equality,
\eql{equality3}{
\tilde{r}_{{\cal B}_{1}}(l)=\tilde{r}_{{\cal B}_{2}}(l)
\mbox{\hspace*{3cm}for 
$1\leq l \leq l^{*}_{{\cal B}_{1}{\cal B}_{2}}$},}
holds within accuracy of the factor $e^{\delta_{\alpha}(l)}$
defined above, as shown in Fig.~2. 
The latter means that 
for $\tw \gg z^{-1}(l^{*}_{{\cal B}_{1}{\cal B}_{2}})$,
the two basins are in quasi-equilibrium. 
Thus more and more basins 
become in quasi-equilibrium with each other as $\tw$ increase.

Let us call the grounp of basins, which are in quasi-equilibrium
with each other
at a given  $\tw$, as a {\it quasi-equilibrium domain} of age $\tw$.
To characterize the growth of quasi-equlibrium domain, we measured
$d_{\rm max}(\tw)$ which is the maximum of the Hamming distance between
pairs of energy minima enclosed in a domain of age $\tw$. 
We found that it is an increasing function of $\tw$.

\subsection{Aging and response to magnetic field}

Lastly, we present a result of our Monte Carlo simulations \cite{TYH}
performed on larger system sizes (but with $J_{ij} = \pm (N-1)^{-1/2}$)
simulating the following well known experimental procedure \cite{experiment}. 
For time $\tw$ after the
temperature quench, the system evolves (ages) under 
zero external magnetic field. Then small magnetic field $h$
is swithched on (at $t=0$) and the induced magnetization $m(t;\tw)$ 
is observed.
We also measure
the spin auto-correlation under zero-magnetic field, 
which is represented, 
by means of the notation in the previous sections, as
$q(t+\tw,\tw)\equiv N^{-1}\sum_{i=1}^{N}\sigma^{\alpha}_{i}\sigma^{\beta}_{i}
p^{\alpha}(t+\tw)p^{\beta}(\tw)$.

\begin{wrapfigure}[12]{r}{\halftext}

  \epsfxsize=6cm \epsfysize=4cm
  \centerline{\epsfbox{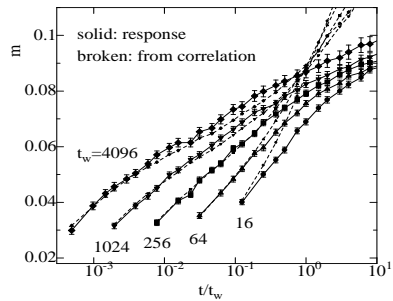}}
  \caption{Response to magnetic field and FDT}
\end{wrapfigure}
In Fig.~3 we plot the data of $m(t;\tw)$ and $h[1-q(t+\tw,\tw)]/T$ 
for $N=512$, $T=0.4$ and $h=0.1$ against $t/\tw$.  
Interestingly, the two quantities coincide with each other, or 
satisfy the flucuation-dissipaton theorm (FDT), 
in the time region $t \ll \tw$. 
The theorem can only be proven theoretically 
assuming the true equilibrium situation 
and the above feature seems highly non-trival.
At present, we speculate that the quasi-equilibrium property 
we discussed before is responsible for this: 
at $t \ll \tw$ fluctuation and dissipation in the system 
are dominated by those in 
the quasi-equilibrium domain 
of age $\tw$, while at later $t$ the system starts 
to invade a bigger domain which causes the breaking of the FDT.

To summarize, we have studied the relaxational modes 
and aging of the SK model
by numerical diagonalization technique and Monte Carlo simulations.
We focused on the charcteristics of the slow modes and found that
the aging proceed by hierarchical growth of quasi-equilibrium domains.

The numerical works have been done on FACOM VPP-500/40
at the Super Computor Center, Institute for Solid State Physics,
University of Tokyo.
This work was supported by Grand-in-Aid for Scientific Research 
from the Ministry of Education, Science and Culture, Japan.
One of the author (HY) was supported by Fellowships of the Japan Society
for the Promotion of Science for Japanese Junior Scientists.



\end{document}